# Modelling the dust content of spiral galaxies: More dust mass vs. enhanced dust grain emissivity

K. M. Dasyra*, E. M. Xilouris†, A. Misiriotis** and N. D. Kylafis**‡

*Max-Planck-Institut für Extraterrestrische Physik, Postfach 1312, D-85741 Garching, Germany
†National Observatory of Athens, I. Metaxa & Vas. Pavlou str., Palaia Penteli, 15236, Athens, Greece
**University of Crete, Physics Department, P.O. Box 2208, 71003 Heraklion, Crete, Greece
‡Foundation for Research and Technology-Hellas, P.O. Box 1527, 71110 Heraklion, Crete, Greece

**Abstract.** We present detailed modelling of the spectral energy distribution (SED) of the spiral galaxies NGC 891, NGC 4013, and NGC 5907 in the far-infrared (FIR) and sub-millimeter (submm) wavelengths. The model takes into account the emission of the diffuse dust component, which is heated by the UV and optical radiation fields produced by the stars, as well as the emission produced locally in star forming HII complexes. The radiative transfer simulations of Xilouris et al. [10] in the optical bands are used to constrain the stellar and dust geometrical parameters, as well as the total amount of dust. We find that the submm emission predicted by our model can not account for the observed fluxes at these wavelengths. We examine two cases, one having more dust embedded in a second thin disk and another allowing for an enhanced submillimeter emissivity of the dust grains. We argue that both cases can equally well reproduce the observed SED. The case of having more dust embedded in a second disk though, is not supported by the near-infrared observations and thus more realistic distributions of the dust (i.e., in spiral arms and clumps) have to be examined in order to better fit the surface brightness of each galaxy.

## INTRODUCTION

Realistic three-dimensional radiative transfer (RT) modelling that was applied to a number of nearby edge-on spiral galaxies [10], [8], [9] has been able to determine the stellar and dust parameters that best fit the optical surface brightness of these galaxies. According to these simulations, a typical spiral galaxy contains $\sim 10^7 M_\odot$ of dust grain material distributed in an exponential disk with a scaleheight about half that of the stars and a scalelength about 1.5 times the stellar one. Furthermore, the extinction law at optical and near-infrared (NIR) wavelengths calculated in these galaxies matches very well the extinction law observed in our Galaxy, indicating common dust properties among spiral galaxies. The opacity of a galactic disk, parametrized by the central face-on optical depth, has values of the order of unity in the B-band.

Complementary to the optical modelling, studies of the far-infrared (FIR) and sub-millimeter (submm) emission of edge-on spiral galaxies have also been carried out. In the studies of Popescu et al. [6], and Misiriotis et al. [5], the far-infrared (FIR) to submm spectral energy distribution (SED) is computed at each point inside a galaxy in a self-consistent way, by calculating the response of the dust grains to the 3D radiation field including the effects of stochastic heating (see also Popescu & Tuffs [7]). In [6] it was found that the inferred from the model submm fluxes were significantly lower than the observed values. To overcome this problem, a second dust disk, with scaleheight matching that of the young stellar population of our Galaxy, was adopted in order to explain the FIR/submm SED. Recent studies though question the validity of the values, widely used so far, for the emissivity of the dust grains in the FIR/submm regime. In particular, a wide range of values (of an order of magnitude) are used for the submm emissivity (see [11], [1] and [4] for reviews of the emissivity values found in the literature).

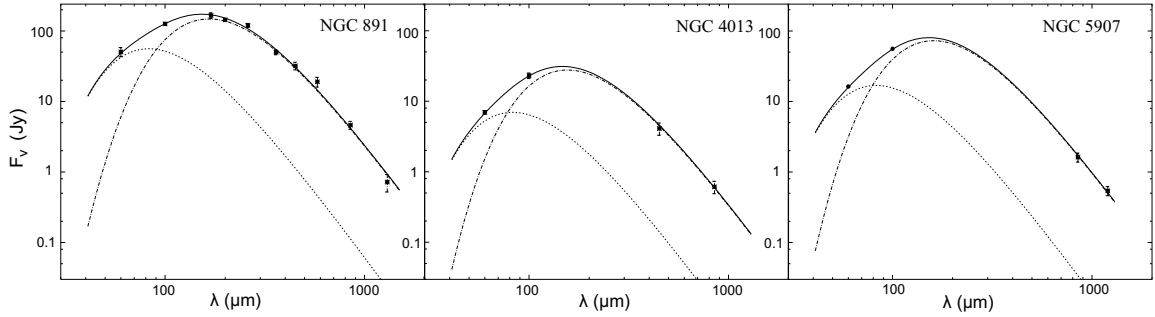

**FIGURE 1.** The observational data of the SED of NGC 891, NGC 4013, and NGC 5907 and our model fits (solid lines). The dotted and dashed lines give the contributions of the HII regions and the diffuse dust respectively. The emissivity is left as a free parameter of the model.

In this study we use a three-dimensional model of the emission of the dust grains in order to fit the FIR/submm SED of three edge-on spiral galaxies (NGC 891, NGC 5907, and NGC 4013) already modelled in the optical wavelengths by Xilouris et al. [10]. Our goal is to examine the two different possibilities (i.e. more dust mass or enhanced submm emissivity).

## THE MODEL

The FIR/submm emission model that we use is based on the principles described in Popescu et al. [6]. In a small volume inside the galaxy, we compute the energy absorbed by the dust, taking into account the three-dimensional distributions of the stars (young and old) and the dust (see also [2]). Having computed the absorbed energy per unit volume it is straightforward to calculate the emission by adopting an emissivity for the dust grains. In this study we examine two cases concerning the values for the dust emissivity. In the first case, the values for the emissivity at different wavelengths are derived by fitting the SED of the galaxy leaving the emissivity as a free parameter. In the second case, the emissivity of the dust grains is set to the values presented in Draine [3]. An additional thin disk of dust grain material associated with the young stellar population in spiral galaxies is added to the model, when the second case is examined, in order to be able to fit the SED.

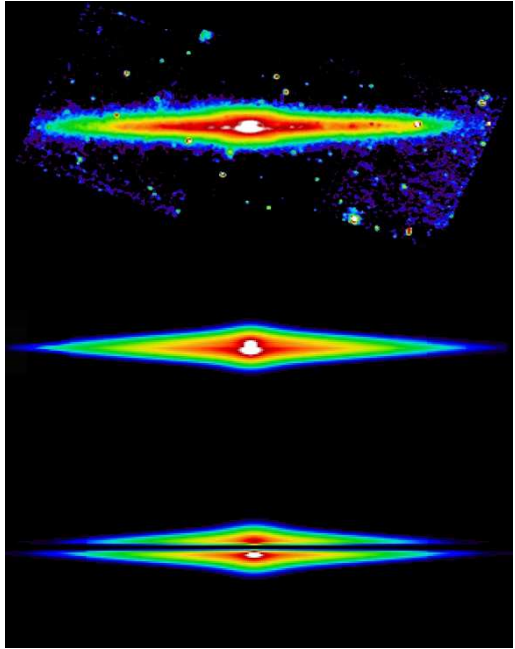

**FIGURE 2.** K-band image of NGC 891 (upper panel), a model with a single dust disk but with an enhanced submm emissivity (middle panel), and a model with an additional dust disk and with a Galactic emissivity (lower panel). The predicted images have been smoothed according to the seeing conditions of the dataset.

Additionally to the diffuse dust component, which is responsible for the bulk of the emission at wavelengths $> 200\mu$m, a component that describes the fraction $F$ of the non-ionizing UV photons, which are locally absorbed by HII regions and re-emitted to FIR wavelengths (quite significant for the $60-200\mu$m wavelength range), first introduced in Popescu et al. [6], is added to the model.

## ENHANCED EMISSIVITY

We first examine the possibility of an enhanced submm emissivity of the dust grains. In order to do this, we leave the emissivity value to vary when fitting the observed SED fluxes. In Figure 1 we show our computed SEDs that best fit the observational data for the three galaxies studied here (namely, NGC 891, NGC 4013 and NGC 5907).

The values derived with the fitting procedure for the emissivity at $850\mu$m are about 4 times the values widely used for our Galaxy (see also [11]).

## MORE DUST MASS

We can reach similarly good fits to the SEDs, as the ones described in the previous section, by adding more dust ($\sim 1-2$ times more) in the form of a second thin disk.

We can see though that such a scenario is not realistic since the K-band modelling, where the extinction effects by dust are marginal, reveals a prominent dust lane when additional dust is embedded in the galaxy. This is shown in Figure 2, where the observed K-band image of NGC 891 (upper panel) is compared with the image created by using a model where dust is distributed in a single dust disk and the emissivity at the FIR/submm wavelengths is enhanced (middle panel) and a model with an additional dust disk and Galactic emissivity (lower panel).

## CONCLUSIONS

We have presented detailed modelling of the FIR/submm SED of three edge-on spiral galaxies (NGC 891, NGC 4013, and NGC 5907). We have demonstrated that the observed SED can be fitted equally well by a model in which the dust (having a submm emissivity $\sim 4$ times the value widely used in the Galaxy) is distributed in a single exponential disk, or by a model where additional dust ($\sim 1-2$ times more) is distributed in a second thinner disk associated with the young stellar population in spiral galaxies and the emissivity is taken to have the Galactic value.

The FIR/submm spectrum of spiral galaxies cannot distinguish between the two models. Thus, we conclude that both scenarios (i.e., enhanced submm emissivity and more dust hidden inside the galaxy) can equally well reproduce the observed SED of a galaxy, while a better representation of the way that dust is distributed in spiral galaxies is necessary.

Future research on the dust distribution within galaxies (e.g. including more realistic cases like spiral arms, dust clumps, etc.) has to be conducted on both edge-on and face-on systems in order to investigate the existence of more dust hidden within the galaxies and thus get a better estimate of the properties of the dust grains.

## REFERENCES

1. Alton, P.B., Xilouris, E., Misiriotis, A., Dasyra, K.M., Dumke, M., 2004, A&A, 425, 109 353, L13
2. Dasyra, K.M., Xilouris, E.M., Misiriotis, A., Kylafis, N.D., 2004, A&A, *submitted*
3. Draine, B.T., 2003, ARA&A, 41, 241
4. Hughes, D., Dunlop, J., Rawlings, S., 1997, MNRAS, 289, 766
5. Misiriotis, A., Popescu, C., Tuffs, R., Kylafis, N., 2001, A&A, 372, 775
6. Popescu, C.C., Misiriotis, A., Kylafis, N.D., Tuffs, R.J., Fischera, J., 2000, A&A, 362, 138
7. Popescu, C.C., Tuffs, R.J., 2004, in "The Spectral Energy Distribution of Gas-Rich Galaxies: Confronting Models with Data", Heidelberg, 4-8 Oct. 2004, eds. C.C. Popescu & R.J. Tuffs, AIP Conf. Ser., in press
8. Xilouris, E.M., Kylafis, N.D., Papamastorakis, J., Paleologou, E., Haerendel, G., 1997, A&A, 325, 135
9. Xilouris, E.M., Alton, P.B., Davies, J.I., Kylafis, N.D., Papamastorakis, J., Trewhella, M., 1998, A&A, 331, 894
10. Xilouris, E.M., Byun, Y., Kylafis, N.D., Paleologou, E., Papamastorakis, J., 1999, A&A, 344, 868
11. Xilouris, E.M., 2004, in "The Spectral Energy Distribution of Gas-Rich Galaxies: Confronting Models with Data", Heidelberg, 4-8 Oct. 2004, eds. C.C. Popescu & R.J. Tuffs, AIP Conf. Ser., in press